\documentclass[twocolumn,11pt]{article}
\usepackage{vmargin}
\setmarginsrb{1.15in}{1.1in}{1.in}{2.0in}{0pt}{0pt}{0pt}{0pt}
\begin{document}

\title{Students' conceptual knowledge of energy and momentum}
\author{Chandralekha Singh and David Rosengrant, University of Pittsburgh}


\date{
We investigate student understanding of energy and momentum concepts at the level of introductory physics by designing and
administering a 25-item multiple choice test and conducting individual interviews.
We find that most students 
have difficulty in qualitatively interpreting basic principles related to energy and momentum and in 
applying them in physical situations. 
The test development process and a summary of results are presented.
}


\maketitle

\vspace*{-.7in}
\section{Introduction}
\vspace{-.15in}

Energy and momentum are two of the most fundamental concepts in physics. 
The goal of this study is to investigate the common difficulties and misconceptions of introductory students pertaining to the
conceptual understanding of energy and momentum. We are interested in understanding the difficulties
students have in interpreting these concepts and in correctly identifying and applying them in different physical
situations. We also want to know the extent to which the difficulties are universal and if there is a correlation
with student preparation (e.g., calculus or algebra background).
Identification of student difficulties can help in designing instructional tools that address them.

To achieve our goal, we designed a research-based 25-item multiple-choice test that explores students' conceptual knowledge of 
energy and momentum~\cite{singh}. The test can be administed as a pre-/post-test to assess the difficulties and misconceptions
prior to instruction, as well as those that remain uncorrected following a particular type
of instructional intervention. The test can be used to compare the understanding of energy and momentum concepts in courses employing different 
instructional designs and strategies. 

Part of the rationale for combining these two concepts is to investigate the extent to which 
students can identify the relevant concept in a particular situation. Students often have difficulty in distinguishing one
from the other. Also, in many physical phenomena, both concepts are simultaneously involved. 
In the introductory courses, energy and momentum are typically taught 
after Newton's laws. 
Widely used research-based tests~\cite{mdt} for assessing force concepts have been designed. They show that students' 
knowledge of force is often fragmented and context-dependent and students have many common misconceptions.

\vspace{-.2in}
\section{Test Design}
\vspace{-.1in}

During the test design, we paid particular attention to the important issues of reliability and validity~\cite{nitko}.
Reliability refers to the relative degree of consistency between testing if the test procedures were repeated for an individual or 
group. Validity refers to the appropriateness of test score interpretation.

The test design began with the development of a test blueprint which provided a comprehensive framework for
planning decisions about the desired test attributes. The degree of specificity in the test plan was a useful guide for
creating items. We tabulated the scope and extent of content covered and the level of cognitive complexity desired.
The energy concepts included 
the work-energy theorem, conservation of mechanical energy, and work done by gravitational and frictional forces.
The momentum concepts included the definition of momentum, impulse-momentum theorem (impulse was defined explicitly in the test),  
and conservation of momentum with examples from
elastic and inelastic collisions. We also planned to evaluate student understanding of the concept of ``system" in various contexts.
We planned to include at least one question that explicitly required the application of both energy and momentum 
concepts.  Energy and momentum questions pertaining to simple harmonic motion, explicit mention of conservative and non-conservative 
forces and center-of-mass reference frame were deliberately excluded because we wanted the test content to be commensurate with the 
curriculum in most calculus- and algebra-based courses for science and engineering majors. 
We simplified Bloom's taxonomy~\cite{bloom}  to classify the cognitive complexity
in three categories: specification of knowledge, interpretation of knowledge and drawing inferences and applying
knowledge to different situations. Then, we determined the specific performance targets for clarifying what is being assessed.
This included identification of desired performance and a description of conditions/contexts under which the performance was
expected to occur. The performance targets and table of content and cognitive complexity were shown to five physics
faculty members at the University of Pittsburgh (Pitt) for review. Modifications were made to the weights assigned to various 
concepts and to the performance targets based upon the feedback 
from the faculty about their appropriateness.

The performance targets were then converted to approximately 50 free-response items. These questions required students
to provide reasoning for their responses. In the Fall of 1998, the free-response items were administered (in groups of
10 or 20) to students in the calculus- and algebra-based courses at Pitt. Often, some students in a class were given
one set of items and others were given another set in order to sample student responses on most of the items. We also
tape-recorded interviews with 10 introductory student volunteers using the think-aloud protocol~\cite{chi}. Forty multiple-choice 
items were then designed using the most frequent student responses for the free-response items and interviews as a guide for making the 
alternative distractor choices. 
Choosing the four distractors to conform to the common difficulties and misconceptions was essential for
increasing the discriminating properties of the items. 

Ten physics facutly members and postdocs were asked to review the multiple-choice questions and
comment on their appropriateness and relevance for introductory physics and to detect ambiguity in item wording.
An item review form was developed to aid the faculty in reviewing the items. The faculty also classified each item on a scale from
very appropriate to least appropriate. Further modifications were made based upon their recommendation. Then, a multiple-choice test
was assembled using 28 items which closely matched the initial table delineating the scope of the content and cognitive complexity.
The same faculty members who reviewed the items were shown the test and slight modifications were made.

The test was administered as a 50-minute post-test to several hundred students in calculus- and algebra-based courses at Pitt in Spring 1999. 
Seven student volunteers who had taken the test were interviewed using the think-aloud protocol.
Item analysis of student responses was performed to judge the quality of each item. 
In addition to the calculation of item difficulty and discrimination~\cite{nitko}, item
analysis included creating a table to count the number of students selecting each
distractor in the upper and lower quartiles. Item analysis was very useful to determine whether individual items
and distractors functioned as expected. Based upon the item analysis and interviews, the test items were modified further.
The number of items in the test was decreased to 25.
Including Spring and Fall 1999, Spring and Fall 2000, and Spring 2001, the test has been administered to more than 3000 students from
approximately 30 calculus- and algebra-based courses in different colleges and universities. 
Some classes administered the test both as a pre-/post-test to assess students' conceptions of
energy and momentum before instruction, the effectiveness of instruction and the effect of pre-test on post-test.
After the administration every semester, an item analysis was performed, a few students were interviewed, and some items were
slightly modified. We have so far conducted 34 one-hour interviews during the test development process. 
For several classes at Pitt, we correlated the performance of students on the test to their
performance on the final exam. We find good correlation which provides evidence for validity. In Fall 2000, the test was 
administered to graduate students enrolled in a teaching
methods class. Their average score was greater than $80\%$ and the reliability coefficient~\cite{nitko} was greater than $0.8$, which provides
further validity to the test.

\vspace{-.15in}
\section{Results and Discussion}
\vspace{-.1in}

The pre-test scores were only slightly higher than random guessing regardless of the class (although
there was a definite pattern and some distractors were more popular than others). We therefore do not calculate the ``normalized gain"
and discuss only the post-test results
of the version administered to 1356 students in Fall 2000. The reliability coefficient (coefficient alpha) for the calculus-based
classes taken together was slightly more than $0.75$ (1170 students) while that for the algebra-based courses was $0.68$ 
(186 students). 
Below, we discuss the post-test results for the calculus-based courses including the honors and active-engagement classes.
The average post-test score was $49.2\%$ (st. dev. $18\%$). 
The item difficulties ranged from $0.2$ to $0.79$. 
The point biserial discrimination~\cite{nitko} (PBD) for individual items ranged from $0.21$ to $0.48$ with only three items with PBD less than
$0.3$. 
The average score on questions focusing mainly on
the energy  concepts (14 total) was $45.7\%$ (st. dev. $20\%$) while that on questions focusing mainly on momentum concepts 
(10 questions) was $55\%$
(st. dev. $22\%$). There was a strong correlation between the performance on the energy questions and momentum questions 
(Pearson~\cite{nitko} correlation $0.54$) indicating that students who performed well on energy concepts typically performed well on momentum concepts.

Although the analysis of variance (ANOVA) shows statistically significant differences between several calculus-based classes
at the level of $\alpha=0.05$,
the effect size $d$ is small ($d<0.35$) for all pairs except those involving an honors class or an active-engagement class.
Also, ANOVA shows a statistically significant difference between the calculus-based (excluding
the honors and active-engagement classes) and algebra-based classes in terms of the overall score, but the effect size
is small. ANOVA on individual items shows that the differences between the calculus- and algebra-based classes are
not statistically significant for 12 of the 25 test items. 

Post-test results show that students lack a coherent understanding of energy and momentum concepts and
have difficulty applying them to different physical situations. 
Many students did not realize that work and energy are scalar quantities and  that momentum is a vector quantity. 
Questions involving work-energy or impulse-momentum theorems were perceived to be more difficult than those involving
their special cases: the conservation of mechanical energy and momentum respectively. However, students had great difficulty
in using the conservation principles appropriately in many situations.
More than $50\%$ of the students did not realize that when a block attached to a string is lifted up a height $h$ at constant 
speed either along an inclined plane or vertically up, the magnitude of work done by gravity and tension are the same in both situations. Many 
(erroneously) reasoned from the fact that a smaller magnitude tension force is required along an inclined plane to conclude that the 
magnitude of work done by tension and gravity are smaller for that case. 

Many students found questions that experts would characterize as similar very different based upon their surface features.
Interviews show that in many physical situations students
knew ``what" but did not know ``why". For example, several interviewed students said that the balls
thrown  from a cliff will reach the ground at the same speed regardless of the angle of projection but
they could not justify their answers based either upon energy or kinematic considerations. 
More than $40\%$ of the students believed that the speed of a person who slides down a slide starting from rest on the top depends on
the mass of the person. Many of them believed that the heavier person has a larger speed at the bottom because a greater downward force
causes a greater acceleration while others felt that the reverse was true because the lighter person is not pressing
down as strongly and has a motion closer to free fall.
$24\%$ of the students
did not realize that if two frictionless slides start and end at the same level then regardless of their actual shape, the speed of
an object released from rest at the top will be the same at the bottom. $11\%$ believed that speed will be larger at
the bottom if the slide had a steeper initial slope.

A large number of students believed that the work done by gravity is path dependent.
More than $27\%$ of the students believed that the work done by gravity on a ball falling from a tower is negative.
Interviews showed that many students did not invoke physics principles to come to this conclusion (e.g., the basic definition
of work) but thought that the work must be negative if the ball is falling in the ``negative y direction".
For an object moving at a constant velocity on a horizontal surface, $46\%$ of students believed that there is a net force
in the direction of motion.
$32\%$ of the students believed that if a bicycle is going up a hill and the cyclist pedals so that the bicycle goes up
the hill at a constant speed, then the total mechanical energy of the cyclist and bicycle is conserved.

Collision problems that were designed to evaluate whether students can identify the appropriate ``system" for
which momentum (and also the kinetic energy for elastic collisions) is conserved show that more than $40\%$ of the
students were confused about it. Many believed that momentum and kinetic energy are conserved for each object.
$24\%$ of the students did not realize that the momentum of an object at a particular instant
depends only on the mass and velocity and not on the acceleration. 
Approximately $10\%$ of the students believed that momentum is force.
Only $28\%$ of the students realized that when identical bullets are fired with the same speed at
two blocks of equal mass (but different material) resting on horizontal frictionless surfaces, the block travels faster if the 
bullet bounces elastically than if it gets embedded inside the block. $36\%$ of students believed that the reason for this is
that the bullet transfers all of its kinetic energy in an inelastic collision while $15\%$ thought it is because 
the bullet that bounces elastically does not impart its momentum to the block (momentum is scalar).

\vspace{-.2in}
\section{Summary}
\vspace{-.1in}

We have designed a research-based test to assess introductory students' conceptual understanding of energy and momentum concepts.
We find that students 
have difficulty in qualitatively interpreting basic principles related to energy and momentum and in applying them in
physical situations. The difficulties and misconceptions were not strongly dependent on student populations or calculus
background, except for honors and active-engagement classes.

\vspace{-.2in}
\section{Acknowledgments}
\vspace{-.1in}

We are very grateful to all the faculty who reviewed the various components of the test at several stages and provided invaluable
feedback. We thank all the faculty who administered the test.

\end{document}